# Correlated states controlled by a tunable van Hove singularity in moiré WSe$_2$

Patrick Knüppel [1*], Jiacheng Zhu [2*], Yiyu Xia [2], Zhengchao Xia [2], Zhongdong Han [1], Yihang Zeng [1], Kenji Watanabe [3], Takashi Taniguchi [3], Jie Shan [1,2,4**], Kin Fai Mak [1,2,4**]

[1] Laboratory of Atomic and Solid-State Physics, Cornell University, Ithaca, NY, USA
[2] School of Applied and Engineering Physics, Cornell University, Ithaca, NY, USA
[3] National Institute for Materials Science, 1-1 Namiki, 305-0044 Tsukuba, Japan
[4] Kavli Institute at Cornell for Nanoscale Science, Ithaca, NY, USA

*These authors contributed equally
**Email: jie.shan@cornell.edu; kinfai.mak@cornell.edu

**Twisted bilayers of transition metal dichalcogenide semiconductors have enabled the discovery of superconductivity[1,2], ferromagnetism[3], correlated insulators[4–6] and a series of new topological phases of matter[7–13]. However, the connection between these electronic phases and the underlying band structure singularities in these materials[14–17] has remained largely unexplored. Here, combining the magnetic circular dichroism and electronic compressibility measurements, we investigate the influence of a van Hove singularity on the correlated phases in bilayer WSe$_2$ with twist angle between 2-3 degrees. We demonstrate stabilizing the Stoner ferromagnetism below moiré lattice filling one and Chern insulators at filling one by tuning the van Hove singularity cross the Fermi level using the electric and magnetic fields. The experimental observations are supported by the continuum model[18,19] band structure calculations. Our results highlight the prospect of engineering the electronic phases by tunable van Hove singularities.**

**Main**
In two dimensions, a divergence in the density of states (DOS), also known as a van Hove singularity (vHS), arises from a saddle point in the electronic band structure. The presence of a vHS near the Fermi level significantly enhances the electron-electron interactions which often lead to electronic instabilities and new phases of matter[20,21]. But typically, in atomic crystals, the Fermi level cannot be easily tuned to cross the vHS. The emergence of the moiré materials[22–26], which support highly tunable electronic band structures and Fermi levels, has provided a new platform to engineer electronic phases by bringing together the vHS and the Fermi level. Among the moiré materials, twisted homo-bilayers of transition metal dichalcogenides (MoTe$_2$ and WSe$_2$) have attracted significant attention recently because they possess flat moiré Chern bands[18,19,27–32] and exhibit a suite of correlated and topological phenomena. These include the integer[8] and fractional Chern insulators[7,9,12,13], integer[11] and fractional quantum spin Hall insulators[10], superconductivity[1,2], ferromagnetism[3] and metal-insulator transitions[4,6,33]. The electronic band structure calculations show a saddle point in the topmost moiré valence band located at the $m$-point of the moiré Brillouin zone. The calculations also show that an electric field perpendicular to the sample plane, which controls the interlayer potential difference[14,18,19], can widely tune the electronic band structure including the location of the vHS[4,6,14,16,17,33]. It has been suggested that proximity of the vHS to the Fermi level can affect the stability of the correlated insulators at integer moiré lattice fillings[14–16,27]. However, a general picture of the effect of the vHS on the symmetry-breaking ground states at generic fillings has remained elusive.

Here we investigate the hole-doped twisted WSe$_2$ (tWSe$_2$) bilayers of small twist angle (2-3 degrees). We demonstrate a ferromagnetic metal below filling one and Chern insulators at filling one by tuning the vHS cross the Fermi level using the electric and magnetic fields. These states are identified by combining the reflective magnetic circular dichroism (MCD)[34] and the exciton sensing measurements[35]. Particularly, the former probes the valley polarization, which is connected to the spin polarization, because of spin-momentum locking—a property inherited from the transition metal dichalcogenide monolayers[36]. The latter probes the sample electronic incompressibility and determines the Chern number of the insulating states through their dispersion with externally applied magnetic field using the Streda formula.

**Phase diagram of tWSe$_2$**

Figure 1a illustrates the dual-gate device structure of tWSe$_2$ employed in this study. The top and bottom gate voltages ($V_{tg}$ and $V_{bg}$) independently control the moiré lattice filling factor ($v$) and the electric field ($E$) perpendicular to the sample plane. A WS$_2$ monolayer, separated from the sample by a thin hexagonal boron nitride (hBN) spacer (about 1 nm thick), is used as the exciton sensor. We focus on 2.7-degree tWSe$_2$ in this study and demonstrate the effect of twist angle at the end. Unless otherwise specified, all results are obtained at sample temperature of $T = 1.6$ K. See Methods for details on the device fabrication, band structure calculations and optical measurements.

Twisted WSe$_2$ bilayers form a honeycomb or triangular moiré lattice depending on the twist angle[37]. The schematic in Fig. 1a illustrates a honeycomb moiré lattice with two sublattices centered at the MX and XM (M = W; X = Se) stacking sites. Figure 1b is the electronic band structure (left panel) and the energy-dependent DOS (right panel) for $E = 0$ reproduced by following the continuum model described in Ref. [18,19]. Only the topmost two moiré valence bands of the K-valley states are illustrated. They carry Chern number $C = +1$. The corresponding moiré bands of the K'-valley states (not shown) carry Chern number $C = -1$. At large electric fields, the two layers are decoupled; the bands become non-topological ($C = 0$) and more dispersive (Extended Data Fig. 1). The DOS shows a vHS that is aligned with the saddle point in the first moiré band and a second vHS from the saddle point in the second moiré band. The hole filling factors at which the Fermi level crosses the first van Hove singularity is about 0.7.

Figure 1c, d are the phase diagrams of tWSe$_2$ as a function of $v$ and $E$. We determine the hole density and $E$ using the applied gate voltages and the gate capacitances. The latter are calibrated using the observed quantum oscillations under high magnetic fields perpendicular to the sample plane (Methods and Extended Data Fig. 2). We benchmark the moiré density, or equivalently, $v = 1$ using the well-established correlated insulator state at half-band filling and determine the twist angle from the moiré density. Figure 1c shows the spectrally averaged reflection contrast of the intralayer exciton resonances of tWSe$_2$.

Representative (raw) reflection spectra at different electric fields and filling factors are included in Extended Data Fig. 3. At a critical field, there is an abrupt change in the reflection signal integrated over the intralayer exciton resonances. Such a change arises from the sensitivity of the intralayer exciton resonance to doping in the corresponding monolayer and has been used to identify the boundary (dashed lines) between the layer-hybridized and layer-polarized regions for small and large fields, respectively. The critical electric field increases with $v$ because the electrostatics requires a larger electric field to fully polarize a higher charge density to one of the layers. The phase diagram is symmetric about $E = 0$ after removing a small build-in field of $E_0 \approx 13$ mV/nm (likely from the device structure asymmetry). These results are fully consistent with the reported phase diagram of tWSe$_2$ (Ref. [1,2]) and tMoTe$_2$ (Ref. [7,9,18,19]).

Figure 1d shows the spectrally averaged reflection contrast of the sensor 2s exciton. The bare reflection spectra as a function of filling factor are shown in Extended Data Fig. 4 for a special case of $E = 0$ (see Methods for details on the spectral analysis). Throughout the measurements, the WS$_2$ sensor remains charge neutral because of the type II band alignment with WSe$_2$ (Ref. [38]). An incompressible state in the sample manifests a blue shift and an enhanced spectral weight of the sensor 2s exciton due to reduced dielectric screening[35,39]. We identify several correlated insulators at commensurate fillings, $v = 1$, 1/3, 1/4 and 1/6, in the layer-hybridized region (the dashed lines are the boundaries from Fig. 1c). These states turn compressible in the layer-polarized region, where the moiré bands are more dispersive, and the correlation effects are expected to be weaker. The apparent asymmetric phase diagram about $E = 0$ is a result of the sensor being closer to the top WSe$_2$ layer.

**Stoner ferromagnetism**
We examine the magnetic properties of tWSe$_2$ by performing the MCD measurements (see Extended Data Fig. 4 for MCD spectra around the intralayer exciton resonances and Methods for extracting the spectrally averaged MCD, which we refer to simply as MCD below). Figure 2a shows MCD as a function of $v$ and $E$ in the absence of magnetic fields. We observe a hot spot of spontaneous MCD around $v = 0.8$ and $E = 0$. It corresponds to a compressible region of the phase diagram (Fig. 1d). The MCD exhibits a clear magnetic hysteresis with a coercive field of about 10 mT (Fig. 2b). As temperature increases, the spontaneous MCD and the magnetic hysteresis gradually weaken and disappear above about 3 K (inset of Fig. 2b). These results support a ferromagnetic metal at the MCD hot spot.

Figure 2c is the MCD under a small out-of-plane magnetic field of $B = 0.5$ T. The result at higher magnetic fields is included in Extended Data Fig. 5. Except in the ferromagnetic region, it increases linearly with magnetic field for small fields, and the small-field MCD is thus proportional to the magnetic susceptibility[22]. The susceptibility is substantially enhanced in the layer-hybridized region below filling one. The susceptibility is also higher in and near the correlated insulators at $v = 1/3$, 1/4 and 1/6. Above filling one, the enhanced susceptibility region disperses with electric field and exhibits an arrowhead-like feature. The dashed lines, at which the MCD drops to 0.03, provide a guide to the eye of the boundary of the enhanced susceptibility region.

To gain insight into the magnetic properties of tWSe$_2$ and the ferromagnetic metal, we perform band structure calculations under varying electric fields. Figure 2d illustrates the electronic DOS as a function of $v$ and $E$ in the experimentally relevant region of the phase diagram. Linecuts at representative electric fields are shown in Extended Data Fig. 1. High DOS is observed in the layer-hybridized region below filling one, where the moiré bands are relatively flat. The vHS is located near $v = 0.7$ for $E = 0$; it continuously shifts towards higher filling factors with reduced DOS as electric field increases. The evolution of the vHS with electric field has been verified by transport measurements[2,4,6] (Extended Data Fig. 7) although the precise location of the vHS in $(v, E)$ is dependent on the twist angle, the choice of the continuum model parameters and can also be affected by the interaction effects that are not accounted for in the single-particle continuum model.

Except in and near the correlated insulators, the measured magnetic susceptibility is well correlated with the calculated electronic DOS. This is expected since the magnetic susceptibility of a Landau Fermi liquid is proportional to the electronic DOS[40]. (In the correlated insulators, the band picture breaks down, and the large susceptibility arises from the local magnetic moments, that is, the spins of the localized holes.) The magnetic susceptibility is substantially enhanced at the vHS. Remarkably, ferromagnetic order is stabilized near the vHS around $E = 0$ where the DOS is the highest. This supports the Stoner mechanism. The Stoner criterion[20], $UD_F > 1$, expressed in terms of the strength of Coulomb repulsion $U$ and the single-particle DOS at the Fermi level $D_F$, provides a qualitative basis for ferromagnetism. The Stoner ferromagnetism in tWSe$_2$ is further supported by the observation of a second ferromagnetic metal phase near but below $v = 3$ around $E = 0$ (Extended Data Fig. 6).

**Chern insulators**
Next, we examine tuning the vHS using the magnetic field and its effect on the correlated insulator at $v = 1$. We focus on the case of $E = 0$. Figure 3a shows the MCD as a function of $v$ and $B$ (lower panel) and the filling dependence of the derivative of MCD at zero field (upper panel). The latter is effectively the magnetic susceptibility and is divergent around $v = 0.8$ for the ferromagnetic metal. Away from the ferromagnetic metal, the MCD increases with magnetic field and saturates till full spin/valley polarization is achieved. Figure 3b displays the field dependence of MCD normalized to its saturation value for several filling factors around $v = 1$. We quantify the saturation field $B_s$ using the value at which the normalized MCD reaches 0.85. The saturation field (dotted line in Fig. 3a) increases rapidly with filling factor, from zero near $v = 0.8$ to several tesla above $v = 1$. We also observe a weak kink in the field dependence of MCD before saturation, which manifests a local minimum followed by a maximum in the derivative of MCD (dashed red line in Fig. 3b). This suggests a metamagnetic transition and the transition field is close to $B_s$.

Figure 3c examines the evolution of the correlated insulators with magnetic field. The lower panel shows the sensor 2s response as a function of $v$ and $B$; the upper panel is a linecut at the highest field of $B = 8.75$ T; the dotted line is the saturation field from Fig. 3a. The insulating states at fractional fillings do not disperse with magnetic field and are likely generalized Wigner crystals[35,41]. The insulating state near $v = 1$ turns into two states above $B_s$. We determine their Chern numbers to be $C = 0$ and 1 based on the Streda formula. For small fields, only the non-topological insulator with $C = 0$ is stable. A potential candidate is a valley-coherent state[14]. The emergence of the Chern insulator with $C = 1$ for fully spin/valley polarized bands is compatible with the spin/valley-contrasting Chern bands in 2.7-degree tWSe$_2$. The coexistence of two types of insulators near $v = 1$ suggests that these competing states are close in the ground state energy. This is district from tMoTe$_2$, where only the Chern insulator is observed regardless of the field (Ref. [7,9]).

We perform the band structure calculations under varying magnetic fields. For simplicity we only consider the Zeeman effect. Figure 3d illustrates the calculated electronic DOS as a function of $v$ and $B$. The vHS with large DOS (dashed lines) splits into two under the field. These features have also been identified in transport studies[1,33]. The vHS that disperses to higher filling factors crosses $v = 1$ at about $B \approx 4$ T. The qualitative agreement between the vHS location and the saturation field suggests the scenario that the vHS tuned to the Fermi level by the magnetic field induces the transition to the spin/valley-polarized state and the emergence of the Chern insulator.

**Twist angle dependence**
Finally, we demonstrate that the observed effect of the vHS near the Fermi level on the correlated states is general in tWSe$_2$ of small twist angles. Figure 4a-d display the evolution of the correlated insulators with magnetic field (upper panel) and the filling dependence of the spontaneous MCD at representative temperatures (lower panel) in tWSe$_2$ with twist angle 1.8 (a), 2.1 (b), 2.3 (c) and 2.5 degrees (d). The black dashed lines denote the expected dispersion for states with Chern number 0 and ±1 near $v = 1$. The results show that Stoner ferromagnetism is stabilized in all but the lowest twist angle sample (1.8°). Chern insulators are absent in the zero-field limit and emerge above the saturation field for all twist angles. In general, decreasing the twist angle lowers the saturation field (arrows) obtained from MCD measurements (not shown). In addition, at twist angle 1.8 degrees, a weak $C = -1$ Chern insulator coexists with the $C = +1$ Chern insulator, whereas only the $C = +1$ Chern insulator is stable for the larger twist angles.

**Conclusions**
In summary, combining observables from the electronic compressibility and magnetization measurements with the continuum model calculations, we develop a general understanding of the electronic phase diagram in tWSe$_2$ and its connection to the vHS in the band structure. Our work bridges the gap[37,42] between the previously explored small[8] and large[4,6,33] twist angle limits. It also provides the basis for vHS engineering of the correlated states in moiré material, such as superconductivity and exciton condensation involving strongly correlated Chern bands[43].

**Methods**
**Device fabrication**
Twisted WSe$_2$ moiré devices were assembled using the layer-by-layer dry transfer method[44], the details of which have been reported in earlier studies[5,7]. In short, thin flakes of hexagonal boron nitride (hBN), few-layer graphite, monolayer WSe$_2$ and monolayer WS$_2$ were exfoliated onto Si/SiO$_2$ wafers. Optical reflection contrast was used to identify the appropriate flake shape and thickness. A thin film of polycarbonate on polydimethylsiloxane (PDMS) was employed as a stamp to pick up the flakes in sequence as shown in Fig. 1a. The complete stack was released onto a Si/SiO$_2$ substrate with prepatterned platinum (Pt) gate electrodes at 180°C. To create the moiré superlattice, a flake of monolayer WSe$_2$ was cut into two parts using an atomic force microscope (AFM) tip, which were stacked with a small relative twist angle of $\theta$.

**Optical reflection contrast (RC) and magnetic circular dichroism (MCD)**
Optical measurements were performed in a closed-cycle cryostat (Attocube, Attodry 2100) with magnetic fields up to 9 T and temperatures down to $T = 1.6$ K. Either a halogen lamp (for 2s sensing and MCD) or a light emitting diode (LED, for MCD) was used as the light source. The input light was spatially filtered by a single-mode fiber and sent into the cryostat in the form of a collimated beam. A low-temperature microscope objective (Attocube, numerical aperture 0.8) was used to focus the light onto the sample. The intensity on the sample was kept below 50 nW/μm$^2$ to avoid perturbing the sample magnetization (no changes in the magnetization were observed by further reducing the incident power by an order of magnitude). The reflected light was collected by the same objective and analyzed by a spectrometer equipped with a liquid-nitrogen-cooled charge coupled device (CCD) array to obtain spectrum $R$. The reflection contrast (RC) spectrum is defined as $(R - R_0)/R_0$, where the reference spectrum $R_0$ was taken at a high doping density, at which the excitonic resonances of the sample are quenched.

The reflective magnetic circular dichroism (MCD) was used to study the magnetic properties of the sample. A linear polarizer and a quarter-wave plate were used to generate the right and left circularly polarized incident light ($\sigma^-$ and $\sigma^+$). The MCD spectrum is defined as $(R^- - R^+)/(R^- + R^+)$, where $R^-$ and $R^+$ are the reflection spectra for the $\sigma^-$ and $\sigma^+$ incident light, respectively.

The reflection contrast spectrum of tWSe$_2$ depends sensitively on the tuning parameters, such as the doping density, displacement and magnetic fields. To account for these changes, we chose to average the total reflectance (by combining the two circular polarizations) over a range of wavelength (725 - 745 nm, or equivalently, 1.66 - 1.71 eV) around the 1s exciton resonance of tWSe$_2$. The averaged reflection contrast is displayed in Fig. 1c. The same wavelength range was used to average the absolute value of the MCD. The averaged MCD is displayed in Fig. 2a. Details of the analysis are illustrated in Extended Data Fig. 4. The magnetic susceptibility was evaluated from the slope of the MCD at small magnetic fields ($|B| \leq 0.5$ T). The magnetic saturation field $B_s$ was defined as the field, at which the MCD reaches 85% of the saturated value. To evaluate the differential susceptibility $\frac{dMCD}{dB}$ (Fig. 3b), we applied a Savitzky–Golay filter to the experimental data before taking a numerical derivative with respect to the magnetic field.

**Determination of the phase boundaries**
The boundary between the layer-polarized and layer-hybridized regions in the $(\nu, E)$ phase space was identified by an abrupt change in the optical refection of the moiré exciton[3]. For

$0.4 \leq \nu \leq 1$, the boundary is sharp, and the electric-field derivative of the reflection displays a clear peak. An example is shown in Extended Data Fig. 3 for $\nu = 1$. The electric-field derivative of the moiré exciton reflection shows a pronounced peak at $E_c = 60$ meV/nm. The dashed green line in Fig. 1c is a cubic spline extrapolation of the boundary electric field to the origin. The dashed black line in Fig. 2c provides a guide to the eye of the boundary of the region with enhanced magnetic susceptibility. We define the boundary at which the MCD (at $B = 0.5$ T) drops to 0.03.

**Optical sensing of compressibility**

Monolayer WS$_2$ was used as the sensor. Changes in the sample compressibility modulate the dielectric environment for the sensor[39], which was probed by its 2s exciton response as demonstrated in Ref. [35,45–47]. The alignment of the sensor and sample valence bands is such that the sensor remains charge neutral for the entire range of $(\nu, E)$ throughout this study. An example is illustrated in Extended Data Fig. 4c for $E = 0$. The reflection contrast spectrum of the sensor around the 2s resonance is displayed as a function of filling factor of the sample (the reference spectrum was acquired when the sensor is electron doped). Both the 2s resonance energy and intensity vary strongly with filling factor of the sample. We used the 2s resonance intensity to represent the sample incompressibility (Fig. 1d). To extract the 2s resonance intensity, we first removed a broad third-order polynomial background for the entire spectral window where both 2s and higher lying excitonic resonances of the sensor are present. We then integrated around the 2s exciton peak (black dots, Extended Data Fig. 4d) over a 2-nm wavelength window for each filling factor.

**Twist angle calibration**

We calibrated the moiré density $n_M$ and the twist angle $\theta$ of WSe$_2$ bilayers using the quantum oscillations observed optically under an out-of-plane magnetic field of 8.8 T. The details for the sample examined in the main text are shown in Extended Data Fig. 2. Extended Data Fig. 2a is the MCD spectrum of the sample near the moiré exciton resonance as a function of gate voltage (or equivalently, hole density). The MCD signal oscillates due to the formation of the Landau levels (LLs). The LL period is determined to be 0.3 V (Extended Data Fig. 2b), from which we deduce a hole density increase of $7 \times 10^{11}$ cm$^{-2}$ per volt. In addition, we identified insulating states through the sensor response as a function of gate voltage and assign the first four most prominent ones to be $\nu = 1/4, 1/3, 1$ and 2. Extended Data Fig. 2c shows these insulating states in filling factor and gate voltage. From the data below filling 1, we determined $n_M = (2.4 \pm 0.1) \times 10^{12}$ cm$^{-2}$ and $\theta = (2.7 \pm 0.1)$ degrees. The results also allow us to determine the hBN thickness for the top and bottom gates: $d_{tg} \approx 18$ nm and $d_{bg} \approx 15$ nm. Using these values we determined the out-of-plane electric field, $E = V_{tg}/2d_{tg} - V_{bg}/2d_{bg} - E_0$, where $V_{tg}$ and $V_{bg}$ denote the top and bottom gate voltages, respectively, and $E_0 \approx 13$ mV/nm is a built-in field likely from the asymmetry of the device in the presence of the sensor layer. We also note that the LL period in gate voltage decreases slightly with increasing filling factor. The nonlinear gating effect likely arises from the non-ohmic contact. We accounted for this effect by using a two-piece linear interpolation of the experimental data points above and below filling factor 1 (Extended Data Fig. 2c).

**Band structure calculations**

We used the continuum model for twisted TMD homobilayers following Ref. [18,19] to compute the single-particle band structure and the density of state of tWSe$_2$. Specifically, the moiré Hamiltonian for the valence band states at the K-valley reads

$$H_K = \begin{pmatrix} \frac{-\hbar k^2}{2m^*} + \Delta_b(\mathbf{r}) + \frac{V_z}{2} & \Delta_T(\mathbf{r}) \\ \Delta_T^\dagger(\mathbf{r}) & \frac{-\hbar k^2}{2m^*} + \Delta_t(\mathbf{r}) - \frac{V_z}{2} \end{pmatrix}. \quad (1)$$

The moiré Hamiltonian for the K'-valley states is a time-reversal copy of $H_K$. Here $\mathbf{k}$ is the momentum, $m^* = 0.43\ m_0$ is the effective hole mass of WSe$_2$ ($m_0$ denoting the free electron mass), $\Delta_{b,t}(\mathbf{r})$ is the bottom (top) layer energy and $\Delta_T(\mathbf{r})$ is the interlayer tunneling amplitude. An interlayer potential difference $V_z$ is also introduced and can be tuned by the out-of-plane electric field. In the long moiré period limit, $\Delta_{b,t}(\mathbf{r})$ and $\Delta_T(\mathbf{r})$ are smooth functions of the spatial position $\mathbf{r}$ in the moiré unit cell and can be approximated as $\Delta_{b,t}(\mathbf{r}) = 2V \sum_{j=1,3,5} \cos(\mathbf{G}_j \mathbf{r} \pm \psi)$ and $\Delta_T(\mathbf{r}) = w(1 + e^{i\mathbf{G}_2 \mathbf{r}} + e^{i\mathbf{G}_3 \mathbf{r}})$ which satisfy all the symmetry constraints of the moiré superlattice. Here $\mathbf{G}_j$ is the reciprocal lattice vectors (lattice constant $a = 3.317$ Å) and $(V, \psi, w) = (9\ \text{meV}, 128°, 18\ \text{meV})$ from Ref. [19] describes the moiré depth, shape and interlayer tunneling strength. The Hamiltonian was cut off at the 5$^{th}$ shell in momentum space. The computed density of states was smoothened using a Gaussian filter with a full width at half maximum of 1 meV. To compare to experiments, we converted the interlayer potential difference to electric field using a dipole moment of 0.26 e·nm[48]. To include the effect of an out-of-plane magnetic field, we added a Zeeman energy shift between the K- and K'-valleys, using a hole g-factor of 10. Specifically, the band structure was first calculated without accounting for the magnetic field. The bands for the K- and K'-valley states were then displaced in energy by 0.58 meV per Tesla and combined to obtain an approximation to the band structure at finite magnetic fields.

**Acknowledgements**
We thank Kaifei Kang, Liguo Ma and Liang Fu for fruitful discussions.

Figures

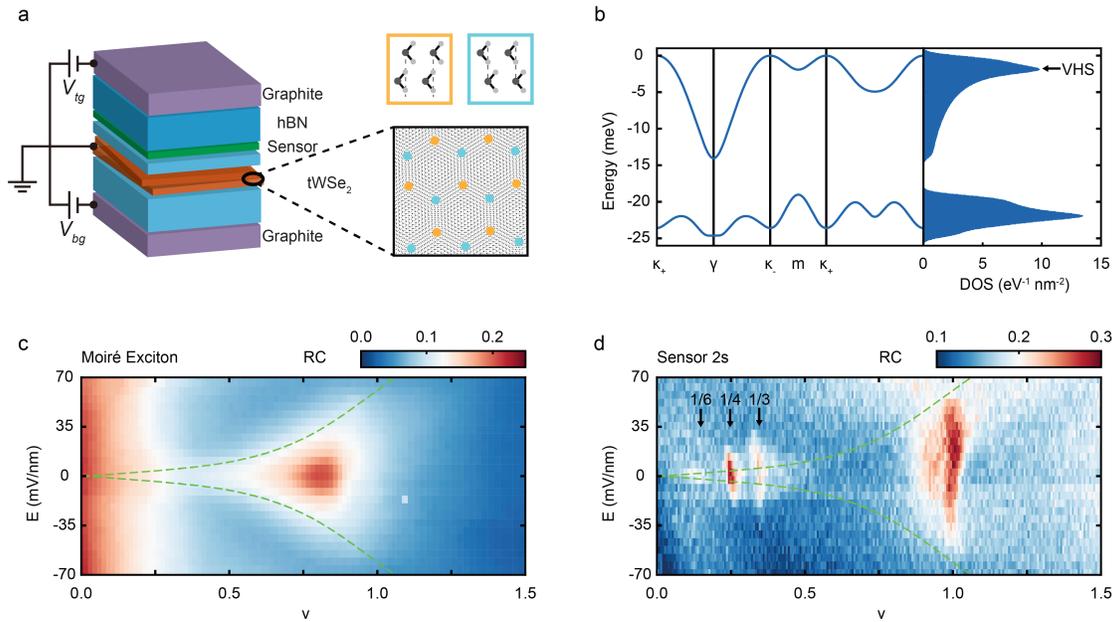

**Figure 1 | Phase diagram of 2.7-degree tWSe$_2$. a,** Left: schematic of a dual-gated tWSe$_2$ device with a WS$_2$ monolayer sensor. Few-layer graphite is used as the gate electrodes, and hexagonal boron nitride (hBN), the gate dielectrics and spacer between the sample and sensor. Gate voltages ($V_{tg}$ and $V_{bg}$) control the moiré lattice filling factor $\nu$ and the out-of-plane electric field $E$. Right: honeycomb moiré lattice consisting of sublattices at the MX (orange) and XM (blue) sites (M = W, black circles; X = Se, grey circles). **b,** Left: continuum model band structure with topmost moiré valence bands of the K-valley state ($E = 0$). Right: density of states (DOS) showing van Hove singularities (vHS). **c,d,** Spectrally integrated reflection contrast (RC) of the sample intralayer exciton (**c**) and the sensor 2s exciton (**d**) as a function of $\nu$ and $E$ at $T = 1.6$ K. The dashed lines separate the layer-hybridized region under low fields from the layer-polarized region under high fields, extracted from **c** as described in the text.

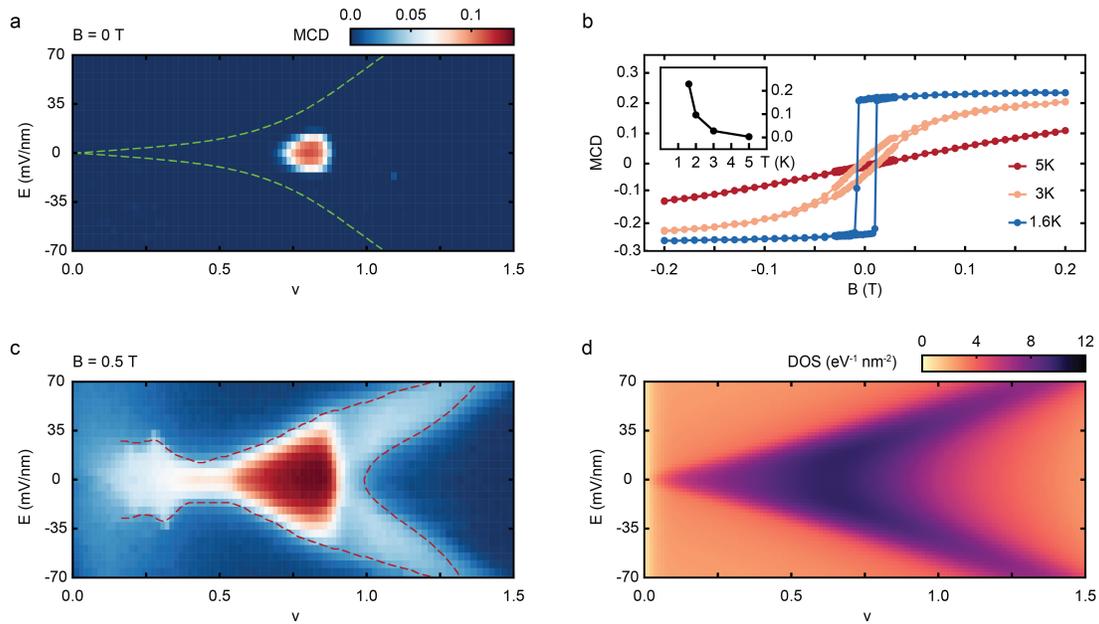

**Figure 2 | Electric-field tuned vHS and Stoner ferromagnetism. a,** Spontaneous MCD as a function of $\nu$ and $E$ at $T = 1.6$ K. The dashed lines (from Fig. **1c**) separate the layer-hybridized and layer-polarized regions. **b,** Magnetic-field dependence of MCD of the hot spot in **a** at representative temperatures. Clear magnetic hysteresis is observed at low temperatures. Inset: temperature dependence of the spontaneous MCD. **c,** Same as **a** under magnetic field $B = 0.5$ T. The dashed lines are a guide to the eye of the boundary of the region with enhanced magnetic susceptibility. **d,** Calculated electronic DOS as a function of $\nu$ and $E$ showing the evolution of the vHS with $E$. All results are for 2.7-degree tWSe$_2$.

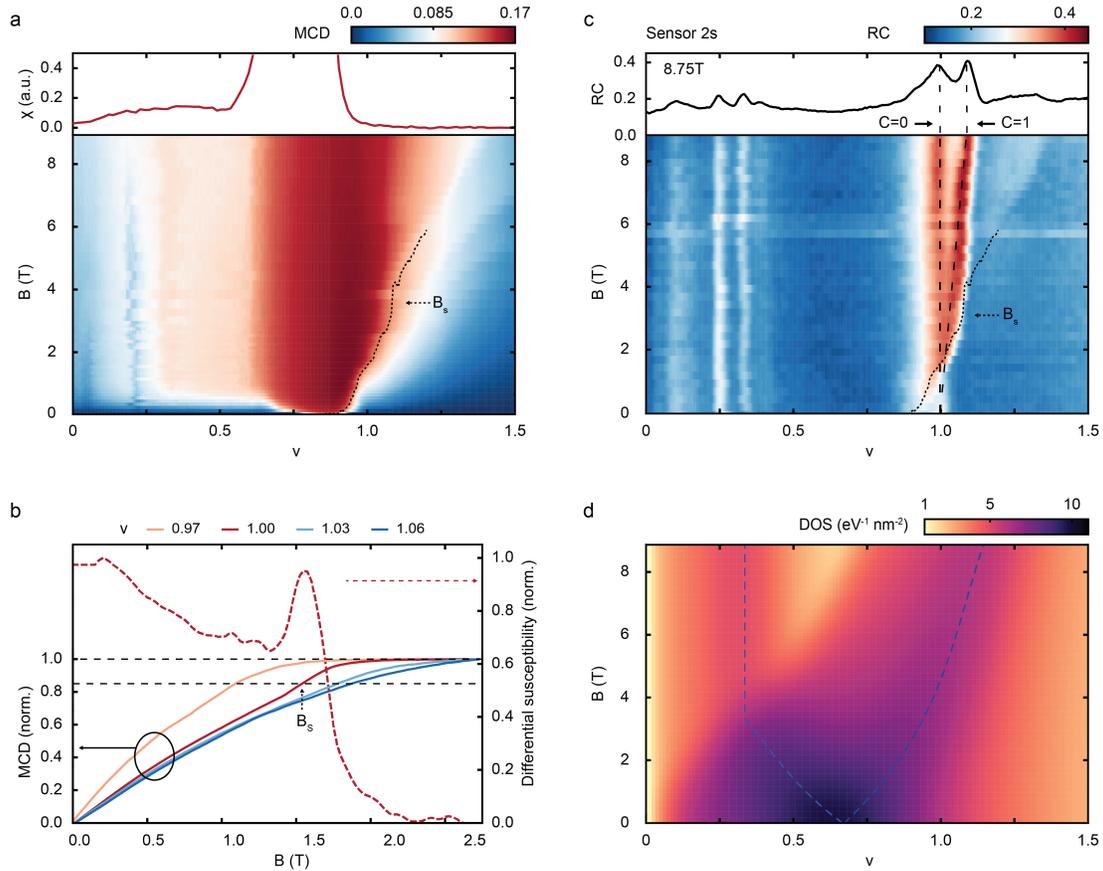

**Figure 3 | Magnetic-field tuned vHS and Chern insulators. a,** MCD as a function of $v$ and $B$ at $T = 1.6$ K and $E = 0$. The dotted line denotes the saturation field $B_s$ at fillings around one. Top: filling dependence of the magnetic susceptibility $\chi$ extracted from the small field MCD. **b,** Magnetic-field dependence of normalized MCD for representative filling factors (solid lines, left axis) and its derivative (differential susceptibility) for $v = 1$ (dashed red line, right axis). The saturation field $B_s$ is defined to be the field at which MCD reaches 85% of the fully saturated value (dashed lines). **c,** Spectrally integrated optical reflection contrast of the sensor 2s exciton as a function of $v$ and $B$ at $T = 1.6$ K. The dotted line denotes the saturation field $B_s$. The dashed lines show the expected dispersion for states with Chern number $C = 0$ and 1 near $v = 1$. Top: linecut at the highest field of $B = 8.75$ T. **d,** Calculated electronic DOS as a function of $v$ and $B$ showing the Zeeman splitting of the vHS traced by the dashed lines. All results are for 2.7-degree tWSe$_2$.

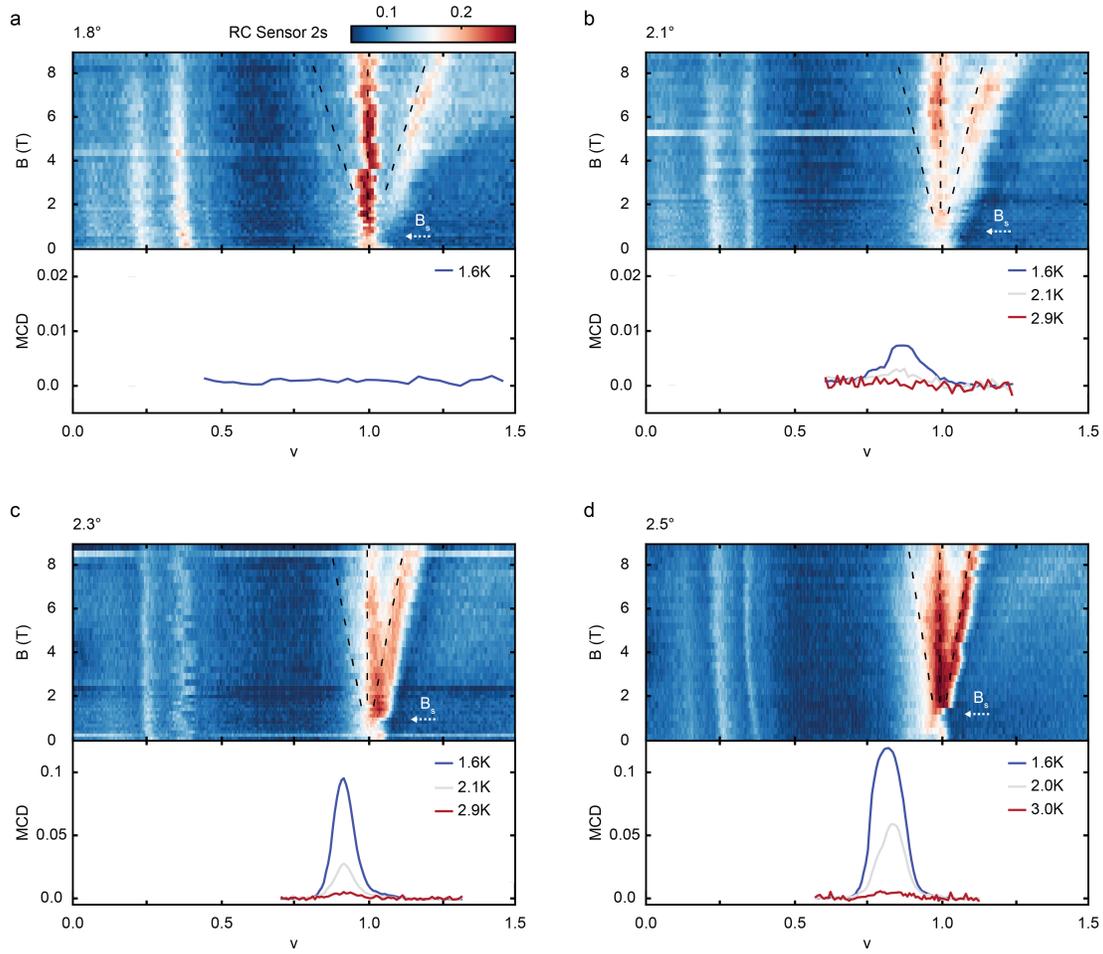

**Figure 4 | Twist angle dependence. a-d,** Top: spectrally integrated optical reflection contrast of the sensor 2s exciton as a function of $v$ and $B$ at $T = 1.6$ K and $E = 0$. The dashed lines show the expected dispersion for states with Chern number $C$ = -1, 0 and 1 near $v$ = 1. The arrow denotes the saturation field and emergence of the Chern insulator(s). Bottom: filling dependence of spontaneous MCD at representative temperatures. The twist angles are 1.8 (**a**), 2.1 (**b**), 2.3 (**c**) and 2.5 degrees (**d**) for WSe$_2$ bilayers. The saturation field decreases with twist angle. In 1.8-degree tWSe$_2$ spontaneous MCD is not observed down to 1.6 K and a weak $C$ = -1 Chern insulator coexists with $C$ = 1 Chern insulator.

**Extended Data Figures**

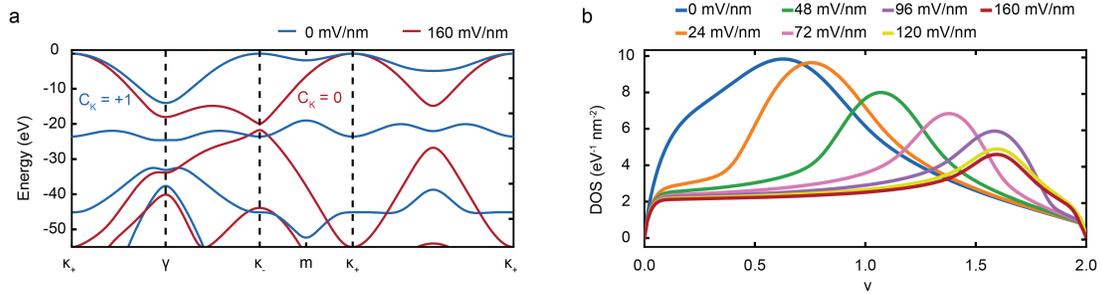

**Extended Data Figure 1 | Electric-field tuned moiré band structure. a,** Continuum model band structure of 2.7-degree tWSe$_2$ at two electric fields. The topmost moiré valence bands of the K-valley state are shown. The bands become more dispersive under a finite electric field. The first moiré band has Chern number 1 and 0 for the small and large electric fields, respectively. **b,** Filling factor dependence of the electronic DOS for different electric fields. The vHS weakens and disperses to higher filling factors with increasing $E$.

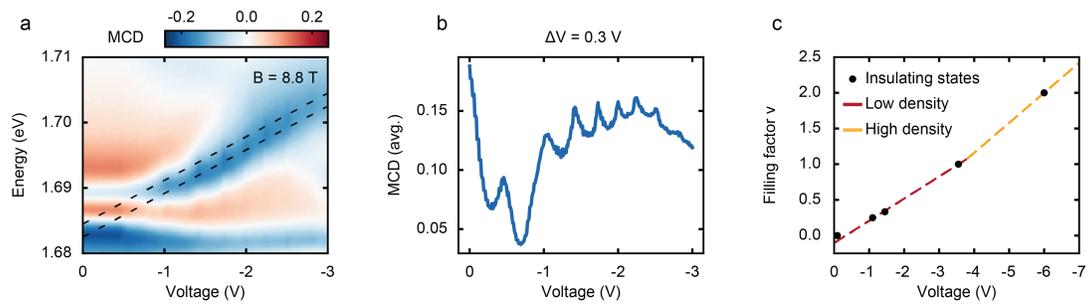

**Extended Data Figure 2 | Twist angle calibration. a,** MCD spectrum of the intralayer exciton of 2.7-degree tWSe$_2$ as a function of (symmetric) gate voltage at $B = 8.8$ T and $T = 1.6$ K. Symmetric gating varies doping but not electric field in the sample. The periodic modulations of MCD show the formation of Landau levels (LLs). **b,** Gate voltage dependence of spectrally averaged MCD over the window between the dashed lines in **a**. The LL spacing is determined to be $\Delta V = 0.3$ V. **c,** Filling factor of the insulating states (identified from 2s sensing) versus gate voltage. A two-piece linear interpolation is used to convert the voltage to filling factor and to determine the moiré density and the twist angle (Methods).

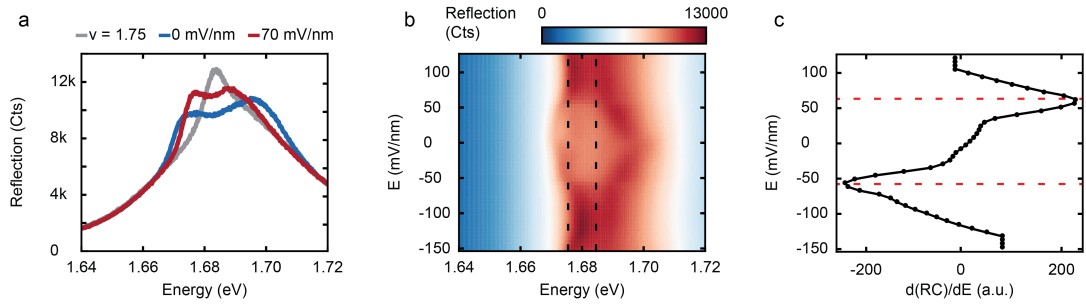

**Extended Data Figure 3 | Determination of the layer-hybridized and layer-polarized phase boundary. a,** Raw optical reflection spectra of the intralayer exciton at $\nu = 1$ and $E = 0$ (blue), $\nu = 1$ and $E = 70$ mV/nm (red), and $\nu = 1.7$ and $E = 0$ (grey). **b,** Raw optical reflection spectrum of the intralayer exciton at $\nu = 1$ as a function of electric field. The dashed lines denote the spectral window employed for integration. **c,** Electric-field derivative of the integrated reflection contrast as a function of electric field. The extrema denoted by the dashed lines define the phase boundary between the layer-hybridized and layer-polarized regions. All results are shown for 2.7-degree tWSe$_2$ at $T = 1.6$ K and $B = 0$.

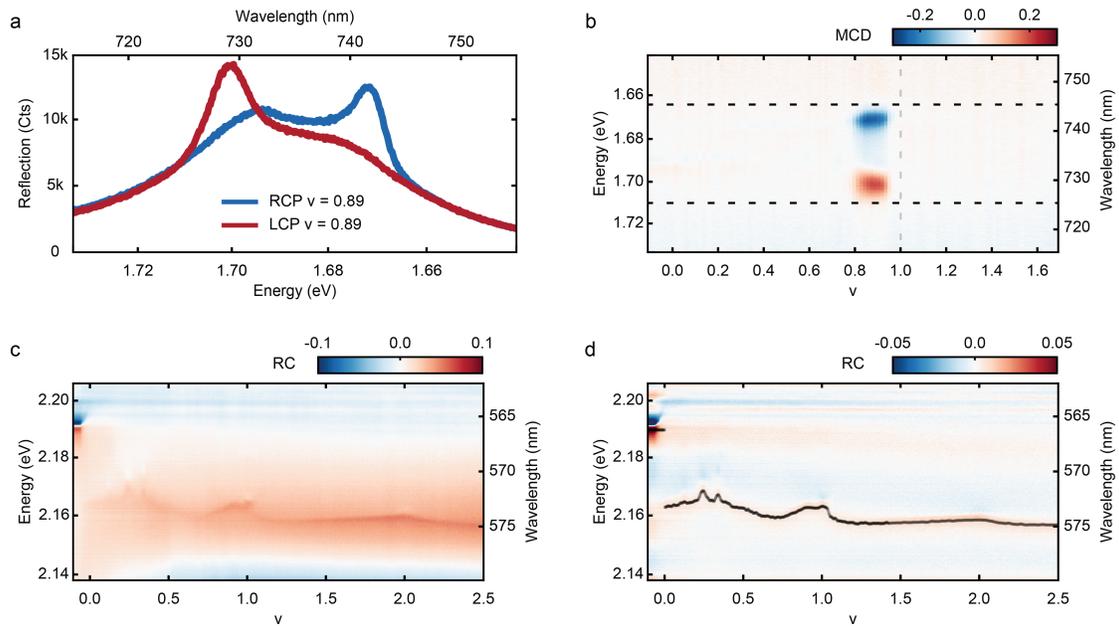

**Extended Data Figure 4 | Analysis of optical reflection spectra. a,** Raw optical reflection spectrum of the intralayer exciton for right and left circularly polarized light ($v = 0.89$, $E = 0$ and $B = 0$). **b,** MCD spectrum as a function of filling factor ($E = 0$ and $B = 0$). The spectrally average MCD is obtained by integrating the absolute value of MCD over the spectral window between the dashed lines. The vertical dashed line marks $v = 1$. **c,** Filling factor dependence of the reflection contrast spectrum of the sensor 2s exciton ($E = 0$ and $B = 0$). **d,** Result in **c** after removing a smooth background (Method) to emphasize the 2s exciton resonance (black line). A spectral window of 2 nm around the 2s resonance is used to obtain the spectrally integrated reflection contrast of the sensor. All results are shown for 2.7-degree t$WSe_2$ at $T = 1.6$ K.

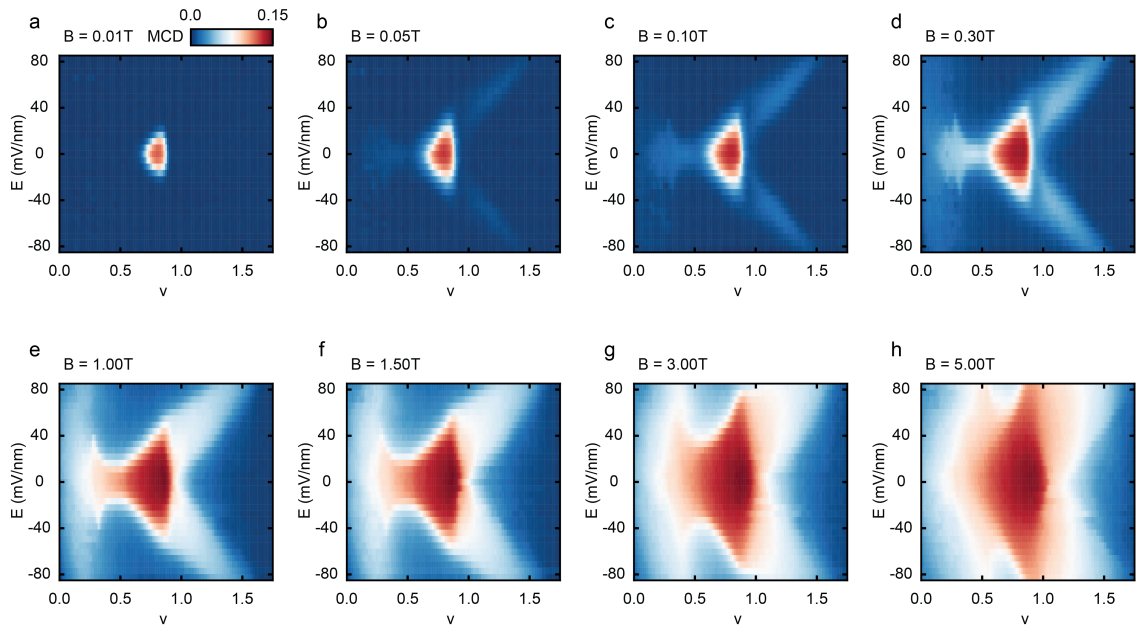

**Extended Data Figure 5 | MCD map. a-h,** MCD as a function of $\nu$ and $E$ under magnetic field of 0 T (**a**), 0.05 T (**b**), 0.1 T (**c**), 0.3 T (**d**), 1.0 T (**e**), 1.5 T (**f**), 3.0 T (**g**) and 5.0 T (**h**). All results are shown for 2.7-degree tWSe$_2$ at $T$ = 1.6 K.

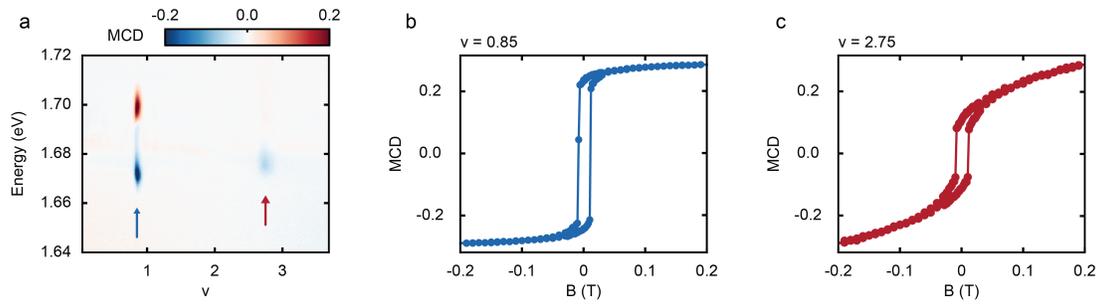

**Extended Data Figure 6 | Stoner ferromagnetism in 2.5-degree tWSe$_2$. a,** Spontaneous MCD spectrum as a function of filling factor. **b,c,** Magnetic-field dependence of MCD at filling $v = 0.85$ (**b**) and $v = 2.75$ (**c**) denoted by the blue and red arrows in **a**, respectively. All results are shown for $T = 1.6$ K and $E = 0$.

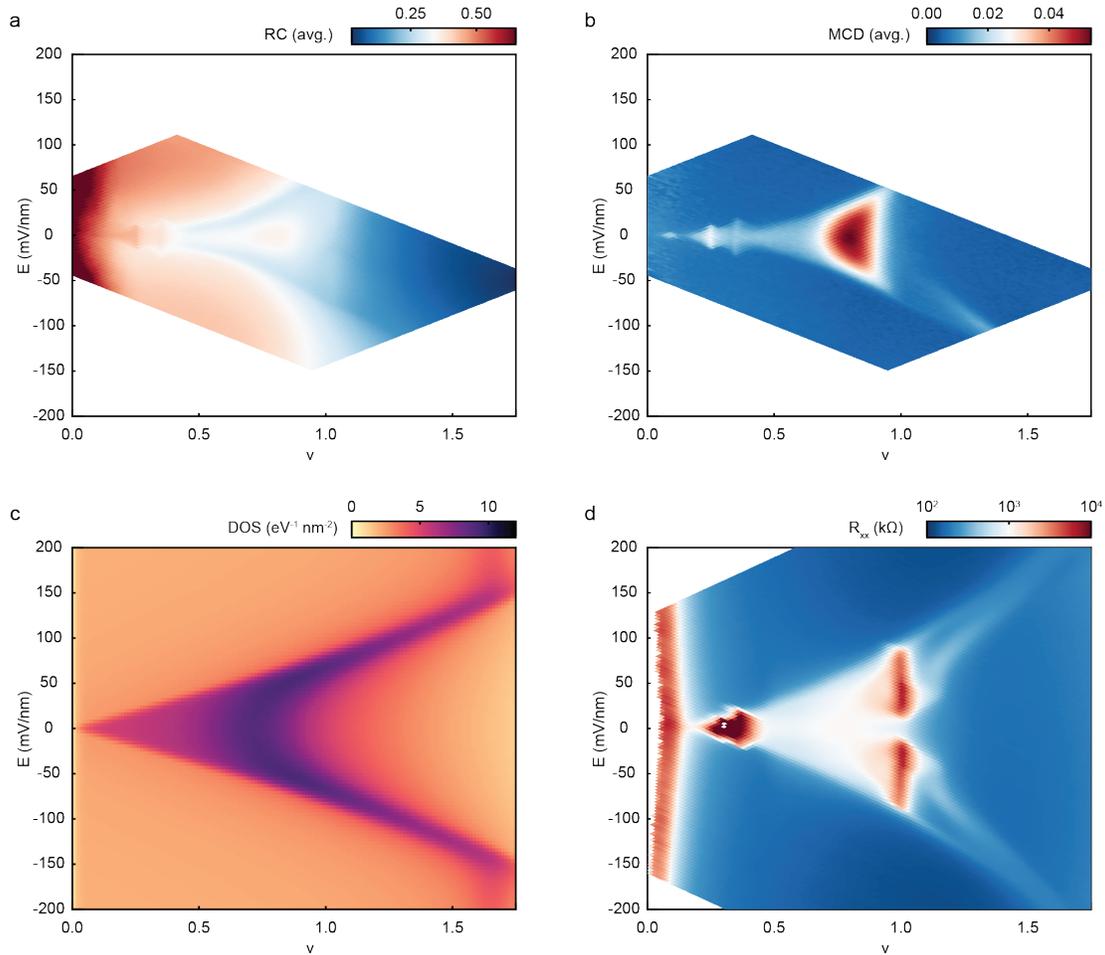

**Extended Data Figure 7 | Comparison of electrical and optical measurements on samples with 3.5° twist angle. a,b,** Spectrally averaged optical reflection contrast for the moiré exciton of the sample (**a**) and the corresponding spectrally averaged MCD (**b**) as a function of $v$ and $E$ at $T = 1.6$ K. **c,** Calculated DOS as a function of $v$ and $E$ showing the evolution of the VHS with $E$. **d,** The longitudinal resistance as a function of $v$ and $E$ at $T = 1.6$ K. All measurements were taken at $B = 0.1$ T. The vHS in the electronic structure is manifested in **d** as a local resistivity maximum. Its location in the electrostatics phase diagram closely matches with that for the enhanced MCD, confirming our picture of vHS-enhanced magnetic response. Note that no metallic ferromagnetism is observed in the 3.5-degree sample due to the weaker correlation effects at larger twist angles.